# An Example of Microwave Diagnosis for Knee Osteophyte by 3D Parallel FD-FDTD Approach

Wenyi Shao, Todd R. McCollough, Arezou Edalati, and William J. McCollough

***Abstract*— A parallel 3-D Frequency Dependent Finite Difference Time Domain (FD-FDTD) method was implemented based on the Single Program Multiple Data (SPMD) technique to analyze the feasibility of microwave diagnosis for the human knees. The parallel algorithm efficiently accelerates the FDTD computation for a large 3-D numerical knee model derived from a real human. Examinations in the frequency domain and time domain were applied to investigate the penetration of the electromagnetic (EM) waves into the knee. Results show that the attenuation of the microwave signal allows for a several-Gigahertz-bandwidth signal to be used for ultra-wide band (UWB) microwave diagnosis. Knee osteophyte detection was undertaken as an example of the knee disease diagnosis to verify this technique. A small abnormal growth in the knee joint was successfully detected by the microwave imaging approach.**

## I. Introduction

Knee problems are very common, and they occur in people of all ages. There are many diseases and types of injuries that can affect the knee. Meniscus, tendons, ligaments, and patella are the most frequently injured joints [1]. The current most commonly used techniques for knee pathologies diagnosis are by nuclear imaging methods [2-4], which are ionizing radiations, and also often expensive. It is important to have an alternative technique to replace/complement to current mainstream methods.

Microwave biomedical imaging is one of the potential alternative technologies and has been increasingly attracting interest and studied [5-9], particularly for early-stage breast cancer detection [10-12]. However, there are not enough studies about feasibility of microwave technique for the knee diseases diagnosis. Since 2009, researchers at the University of Calgary have explored applying microwave imaging to detect meniscal tears and ligament-tendon ruptures [13-15]. In their investigation, a simplified knee model containing selected tissues of the knee joints and canola oil (serving as the coupling liquid) was employed without the consideration of the effects from muscle and skin. Their results showed success of detecting the existence of a lesion in menisci, ligament, or tendon, but are not very sufficient to demonstrate the feasibility of knee microwave imaging because significant reflections and losses due to the muscle and skin were not accounted for. Further, the diseases studied are very close to the surface of the knee, which does not require a deep penetration of microwave signals. Other researchers, for example in [16], have studied using chirp pulse microwave to visualize the physiological or biochemical changes in arms and legs. A 2-D model derived from an MRI scan was applied in their investigation and the frequencies employed ranged from 2-3 GHz, which is insufficient to provide a high-quality image with a good resolution. Therefore, further assessments must be made if diseases deep in the knee are to be detected by microwave method, or a full image of the knee with high resolution is desired based on the electrical-property reconstruction by inverse methods.

Finite difference time domain (FDTD) method has been widely employed to analyze electromagnetic (EM) medical problems [17-19]. In this paper, we demonstrate the feasibility of the broad band microwave method for human knee diagnosis using a 3-D parallel FD-FDTD. The knee phantom was extracted from a full human body model deriving from high-resolution MRI scans of healthy volunteers [20] and was mathematically modeled by Gabriel's method [21-23]. The knee phantom employed in our study has 10 types of tissues including skin and muscle. Due to the dispersive properties of human tissues, FD-FDTD was applied in our simulation, and a parallel algorithm based on SPMD multicore was used to accelerate the computation. Frequency domain and time domain analyses were explored to test the penetration of the microwave signal in the knee. Our frequency-domain analysis went up to 9 GHz that has not been concentratively studied in previous biomedical microwave imaging. As an example, we attempted to detect and locate the position of an osteophyte that occurs in the most common place in the knee using the data obtained from our FDTD simulation. The reconstructed image successfully shows the place of disease.

In Section II of this paper, we elaborate on our parallel FDTD implementation including how the 3-D model was split and the data was distributed. FD-FDTD is also discussed in this section. Section III discusses the frequency and time domain responses of the knee when it is illuminated by a broad band pulse in the form of plane wave or point source. In Section IV, osteophyte detection in the knee using the method previously applied for breast cancer detection is applied. Conclusions and prospects for future studies are in the last section of this paper.

W. Shao is with EMAI LLC, laurel, MD 20723 USA; T. McCollough and W. McCollough are with Ellumen Inc., Silver Spring, MD 20910 USA; A. Edalati is with Varex Imaging Corp.

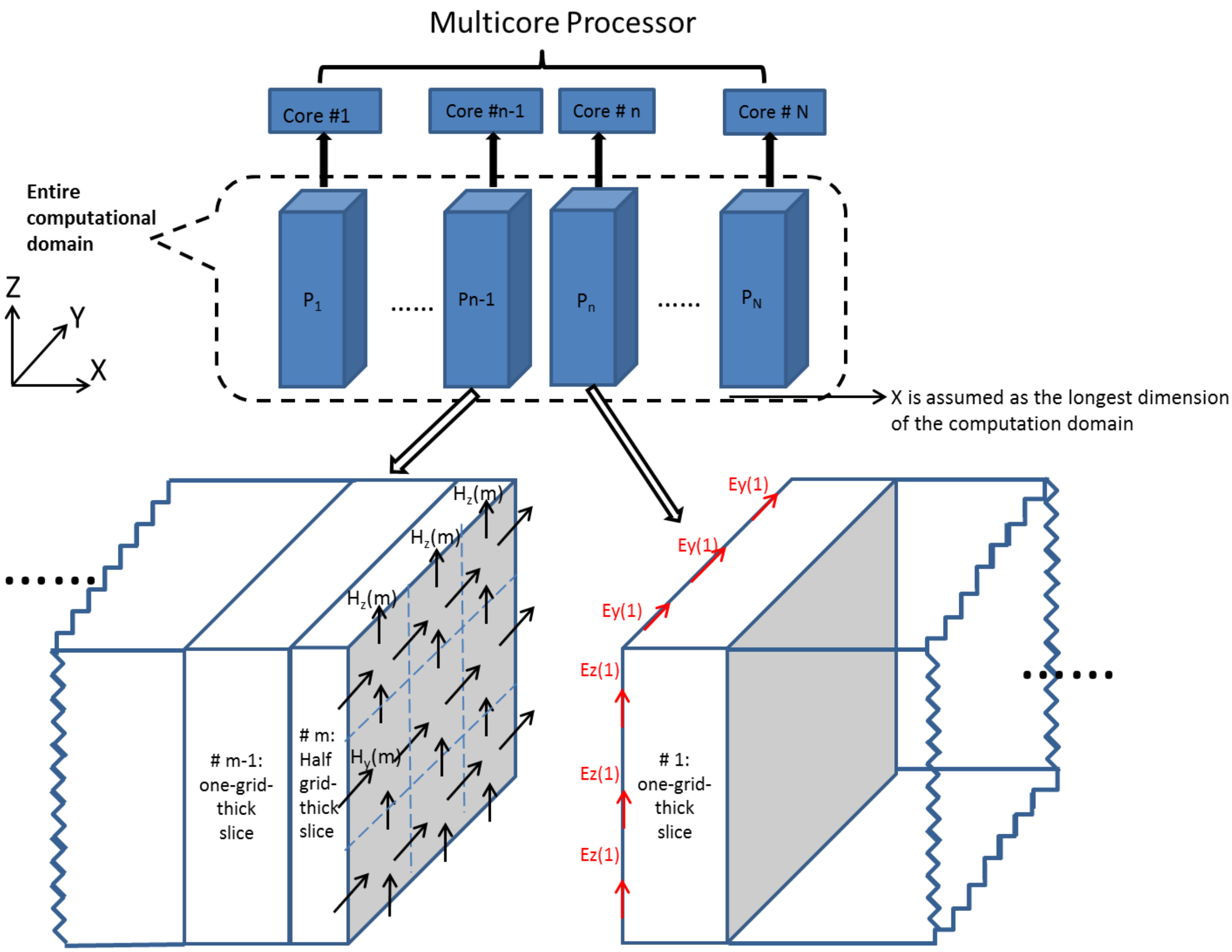

Figure 1. The architecture of a distributed model on a multi-processor computer.

## II. FDTD Implementation

According to Maxwell's curl equations, the electric field and magnetic field are updated by the discrete equations:

$$
\begin{aligned}
E^{n+1} &= CA \cdot E^{n} + CB \cdot [\nabla \times H]^{n+\frac{1}{2}} \\
H^{n+\frac{1}{2}} &= DA \cdot\ H^{n-\frac{1}{2}} - DB \cdot [\nabla \times E]^{n}
\end{aligned} \tag{1}
$$

where $E$ can be $E_x$, $E_y$, or $E_z$, and $H$ can be $H_x$, $H_y$, or $H_z$. *CA*, *CB*, *DA*, and *DB* are coefficients determined by the local dielectric parameters and the selected time-step length. In many cases, magnetic conductivity is thought to be zero, leading to the coefficient of $H^{n-\frac{1}{2}}$ equal to unity. We may either distribute the dielectric parameters, or the coefficients *CA*, *CB*, *DB* in the model initialization step.

### A. Distributed Data and Parallel FDTD

Fig. 1 shows the way the computational domain is split in one dimension along the X axis, assuming it is the longest dimension of the model and there are *N* cores, each processing a subdomain of the entire model. In this manner of partition, the cutting interface between subdomains would have the smallest area, meaning that data delivered from one subdomain to its neighbor are reduced, resulting in economical communication time. Our parallel FDTD code is written in MATLAB language using the SPMD approach [24]. MATLAB distributes the model into *N* subdomains in equal size by default for efficiency (one can also manually distribute the model unevenly). Each subdomain can be assumed to be divided into slices (*m* slices, in Figure 1) along the X direction and each slice is a Y-Z plane with one-grid thick. In a Yee cell, ***E*** fields are along the edges of the cube while ***H*** fields locate in the center of each face of the cube and perpendicular to the face. We physically located $E_z$ and $E_y$ on the subdomain-cross boundary onto the left face of the *1*$^{th}$ slice in the *n*$^{th}$ subdomain as shown in red arrows, except for the *N*$^{th}$ subdomain. Thus, the *m*$^{th}$ slice of each subdomain (except for *N*$^{th}$ subdomain) can be thought as half-grid thick, and its right face contains $H_z$ and $H_y$, shown in black arrows in Fig. 1.

According to Equation (1), the electric field is determined by its last time step and its surrounding 4 magnetic field components at last half time step (and vice versa). In Fig. 1, updating $H_z$ for the *m*$^{th}$ slice in the *n-1*$^{th}$ subdomain requires $E_y$ for the 1$^{st}$ slice in the *n*$^{th}$ subdomain (and vice versa); updating $H_y$ for the *m*$^{th}$ slice in the *n-1*$^{th}$ subdomain requires $E_z$ for the 1st slice in the $n^{th}$ subdomain (and vice versa). Message Passing Interface (MPI) technique, usually containing hundreds of lines of code, is often used to deliver data between subdomains in parallel FDTD [25, 26]. Using the MATLAB parallel computing toolbox, one can use the “labSendReceive” function to transfer data without writing complicated MPI programs. Note that the $1^{th}$ core (most left) does not receive data, but only sends; the most right core does not send data, but only receives. H field data are exclusively transferred from the left to the right subdomain, while E field data are exclusively transferred from the right to the left subdomain, and the transferred data are a 2-D matrix (a slice). Updating $E_x$ for the $m^{th}$ slice in the $n\text{-}1^{th}$ subdomain, and $H_x$ for the $1^{st}$ slice in the $n^{th}$ subdomain do not need communication across the subdomains. The most left (left surface of the 1st subdomain) and the most right (right surface of $n^{th}$ subdomain), as well as top, bottom, front, and back surfaces for each subdomain are normal absorbing boundaries such as PML.

### B. *Curve Fitting and Frequency-dependent FDTD*

Based on the anatomical model off of Christ [20], we mapped the tissues to equivalent tissues used in Gabriel’s 4-pole-Cole-Cole model [27-29] based on measurement data. The anatomical model from Christ consists of more than 80 tissue types whereas the number of tissues in Gabriel’s 4-pole-Cole-Cole model is 45. The knee anatomical model from Christ consists of 15 tissues whereas the number of tissues mapped to Gabriel’s 4-pole-Cole-Cole model is 10 with an additional mapping to air. We noticed a potential issue with our model as tendons and ligaments are usually thought to be anisotropic in structure due to collagen fibers. However, a recent study has shown that anisotropy is unseen in microwave frequency [13]. Therefore we assume the phantom we used with no consideration of anisotropy is still valid for our study.

Due to the dispersive nature of human tissues, a frequency dependent technique in the FDTD computation allows us to accurately model the wave propagation in dispersive media. Gabriel’s 4-pole-Cole-Cole model is valid in a very wide bandwidth from 10 Hz to 20 GHz. The 4-pole-Cole-Cole equation contains a $(jw)^{1-\alpha}$ term (where α is a decimal) which causes a fractional-order differentiator involved in the recursive FDTD formula. There have been several approaches to carry this out [18-20], but these methods are usually complicated and time consuming in simulation. Therefore, we curve fitted the 4-pole-Cole-Cole parameters into one-pole Debye dispersion parameters that are more readily incorporated into FD-FDTD. The one-pole Debye equation is

$$\tilde{\varepsilon}(\omega) = \varepsilon_\infty + \frac{\varepsilon_s - \varepsilon_\infty}{1 + j\omega\tau} - j\frac{\sigma_s}{\omega\varepsilon_0} \quad (2)$$

where $\varepsilon_\infty$ is the permittivity value at infinite high frequency, $\varepsilon_s$ is the static, low frequency permittivity, $\tau$ is the characteristic relaxation time, $\sigma_s$ is the static conductivity, and $\varepsilon_0$ is the permittivity in vacuum. We performed curve fitting by minimizing a function in the least squares sense between the 4-pole-Cole-Cole model and the Debye model [30]. A trust-region-reflective algorithm [31] was used in our curve fitting process. Fitted parameters in the Debye equation have four variables $\varepsilon_\infty$, $\varepsilon_s$, $\sigma_s$, $\tau$, which were then used to compute the dielectric constants and conductivities in the frequency range from 1 to 10 GHz. Fig. 2 shows the comparison of the dielectric constant and conductivity versus the frequency between the 1-pole Debye and 4-pole Cole-Cole models for two tissues in the knee model: infiltrated bone marrow and muscle. The accuracy of our curve fitting was found to be acceptable between 1 and 10 GHz. Fig. 3 shows the dielectric parameters for all 10 types of tissues in our knee model using one-pole Debye equation.

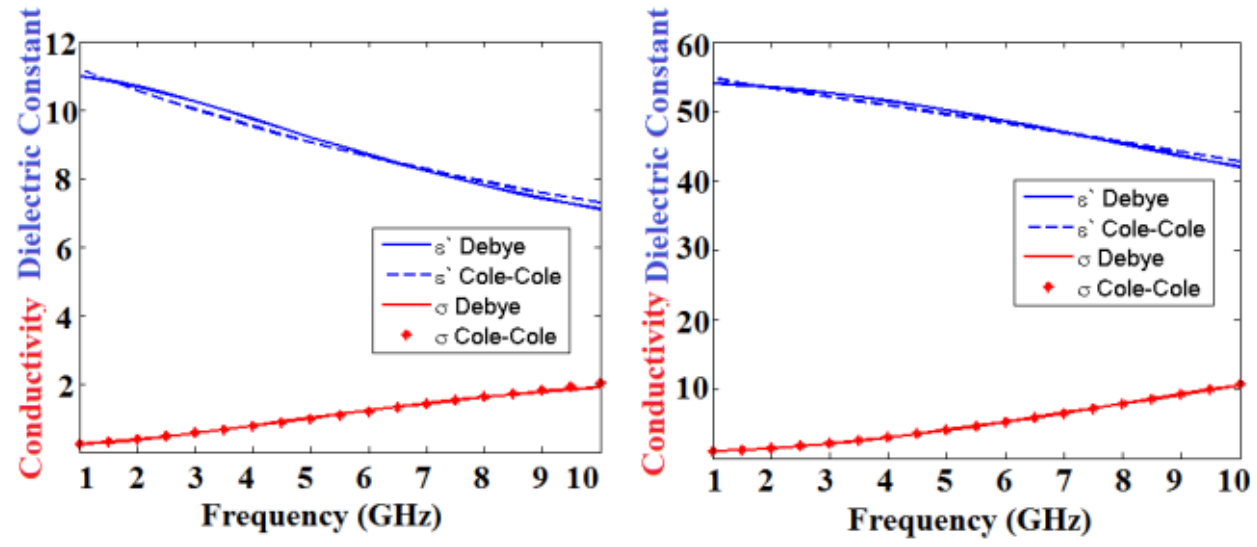

Figure 2 Dielectric constant (blue) and conductivity (red) over the frequency 1-10 GHz for 4-pole-Cole-Cole equation and 1-pole-Debye equation curve fit for bone marrow tissue (left), and muscle tissue (right).

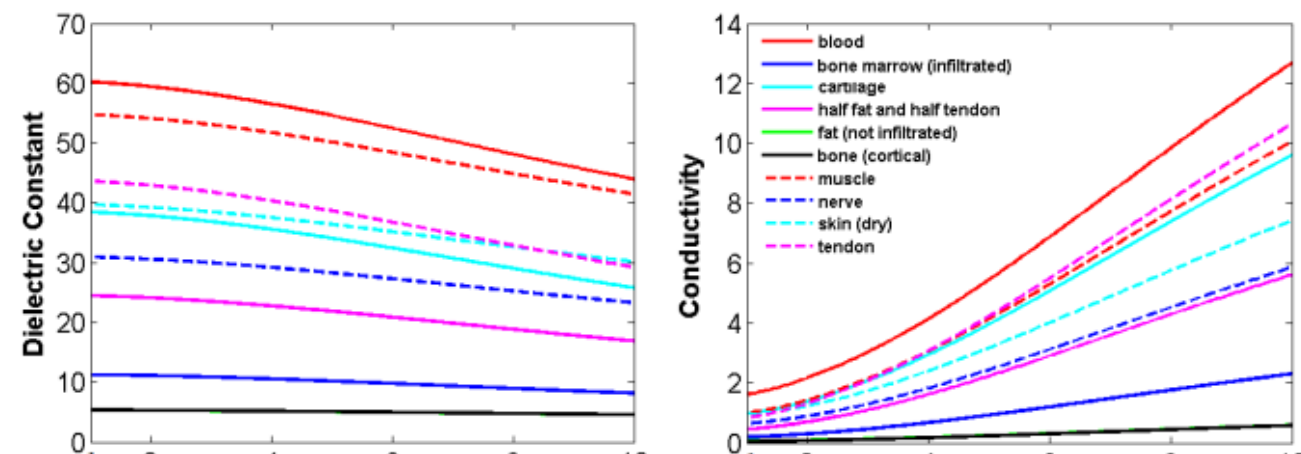

Figure 3. The dielectric parameters of the tissue in the knee model: (a) dielectric constant; and (b) conductivity.

Equation (2) is incorporated into the constitutive equation $D = \tilde{\varepsilon}(\omega)E$ to compute the $E$ field by the updated electric displacement field $D$. Applying frequency-to-time conversion $j\omega = \partial/\partial t$, discretizing the differential equation, and after some arrangements, one may write out the discrete equation in the form of

$$E^{n+1} = A_0 D^{n+1} + A_1 D^n + A_2 D^{n-1} - B_1 E^n - B_2 E^{n-1} \quad (3)$$

where, $A_0$, $A_1$, $A_2$, $B_1$, and $B_2$ are coefficients, which contain the four curve-fitted parameters created before. Equation (3) denotes that updating the $n\text{+}1^{th}$ step $E$ field requires the information of the $n\text{+}1^{th}$ $D$ field, $n^{th}$ $D$ field, $n\text{-}1^{th}$ $D$ field, $n^{th}$ $E$ field, and $n\text{-}1^{th}$ $E$ field, implying that more memory is required in FD-FDTD. In order to have $D$ field to compute $E$ field, one may use

$$D^{n+1} = D^n + \Delta t \cdot [\nabla \times H]^{n+\frac{1}{2}} \quad (4)$$

where $\Delta t$ represents time-step length. Accompanied with the second formula in equation (1), the sequence for field updating in FD-FDTD can be summarized in $E^n \rightarrow H^{n+\frac{1}{2}} \rightarrow D^{n+1} \rightarrow E^{n+1}$.

To parallelize frequency-dependent FDTD, one can either distribute four dielectric parameters$(\varepsilon_\infty, \varepsilon_s, \sigma_s, \tau)$, or compute coefficients *DB*, $A_0$, $A_1$, $A_2$, $B_1$, and $B_2$ (*DA* is thought to be one) on the client side and distribute these coefficients. The former method reduces memory but has low efficiency due to repeating coefficients in computation. If memory is not an issue, the latter method is preferred which was adopted in our simulations. Note that the positions of *D* field are exactly the same to *E* field. Therefore, the architecture of distribution of *D* field is identical to E field introduced in Section II. In addition, only *E*-field slices and *H*-field slices are delivered between subdomains. There is no *D*-field data transfer.

## III. Electromagnetic Computation For The Knee

In this section, frequency-domain investigation and time-domain investigation are elaborated respectively. The plane wave excitation is used in the frequency domain analysis, and point source is applied in the time domain analysis.

### *A. Frequency-domain Analysis*

The numerical knee model implemented in our investigation is a $X \times Y \times Z = 800 \times 640 \times 640$ rectangular structure with a grid size of $\Delta x = \Delta y = \Delta z = 0.25\,mm$. Fig. 4 shows our 2-D cuts from the 3-D model. The model was cut by PMC in the X-Z planes on the left and right and by PEC in the X-Y planes on the top and bottom displayed in Fig 4 (b). A plane wave excited in the front of the knee is framed within this PMC-PEC bounded scope, propagating from the front to the back (+Z direction) in Fig. 4(a) and linearly polarized in the X direction. Thus, the waves are forced to pass through the knee instead of passing by. The entire model is assumed to be immersed in a coupling medium [32-34] with dielectric constant $\varepsilon_r = 10$. It took 30 hours to accomplish 12,000 time steps in the FD-FDTD simulation by an eight-core parallel acceleration on our Intel Xeon 3.33GHz computer (8 core).

The electric field in a transverse plane when X=100 mm, and a vertical plane when Z=150 mm which is behind the knee (both are shown in green-dashed lines in Fig. 4 (a)), was recorded during the entire simulation in time domain. The time domain data were converted into the frequency domain at each point in the plane by a discrete Fourier Transform.

Fig. 5(a-d) shows the electric field at 3, 5, 7, and 9 GHz respectively in the transverse observation plane. As expected, the penetration of the field reduces as the frequency increases. At 3 GHz the electric field inside the knee is mostly above -60dB, whereas at 7 GHz and 9GHz the field inside the knee can be as low as -80 dB or less. It was also observed that the field in the center part of the knee does not have large differences from 6 GHz to 9 GHz (only 7 GHz and 9 GHz are shown for briefness). The reason for this phenomenon is likely because the dominant

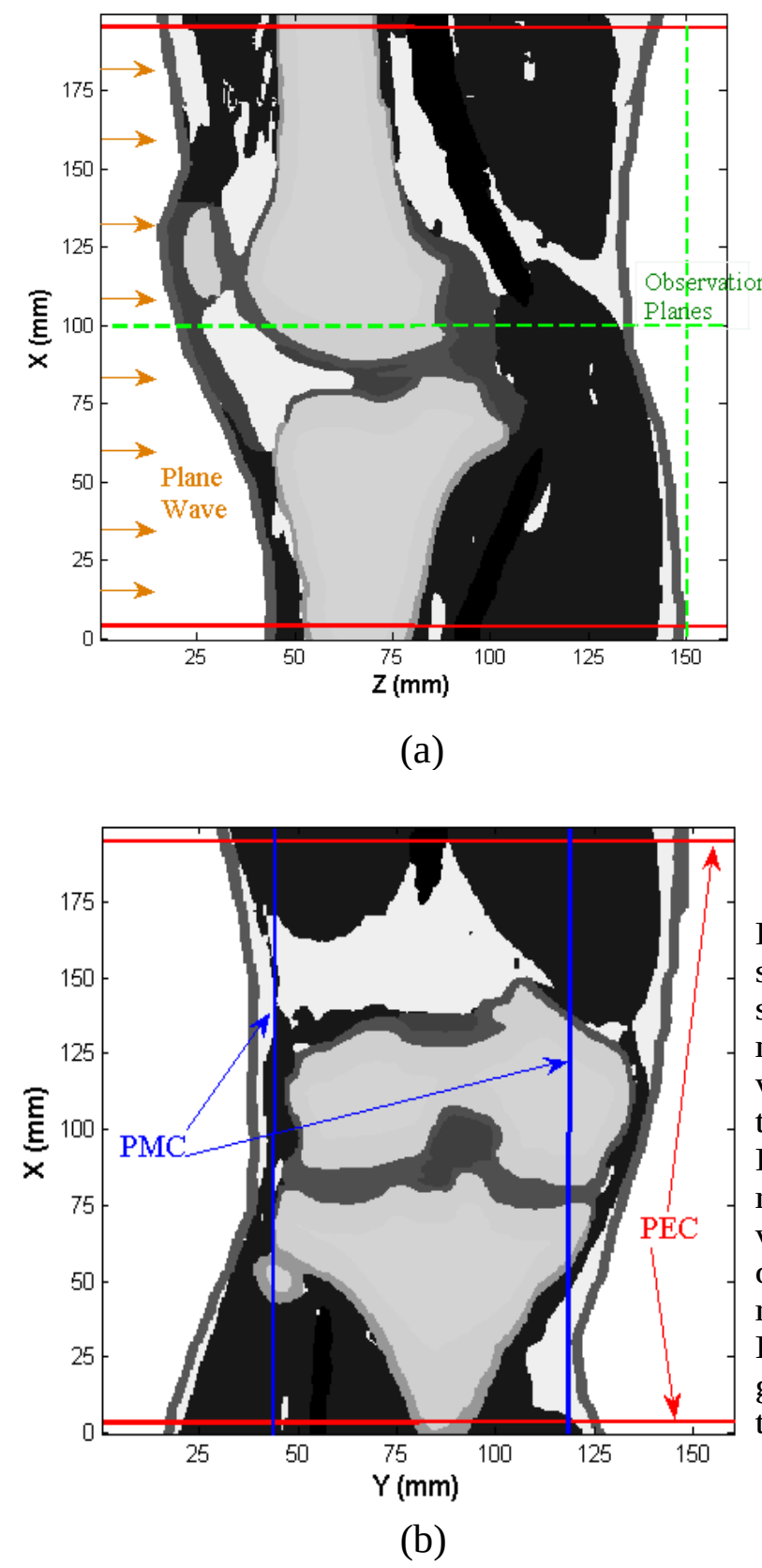

Figure 4. The X-Z-plane slice (a) and X-Y-plane slice (b) of the 3-D knee model. The knee model was cut by PEC on the top and bottom, and by PMC on the left and right. The plane wave was illuminated in front of the knee and was restricted within the PEC-PMC frame. The green dashed lines show the observation planes.

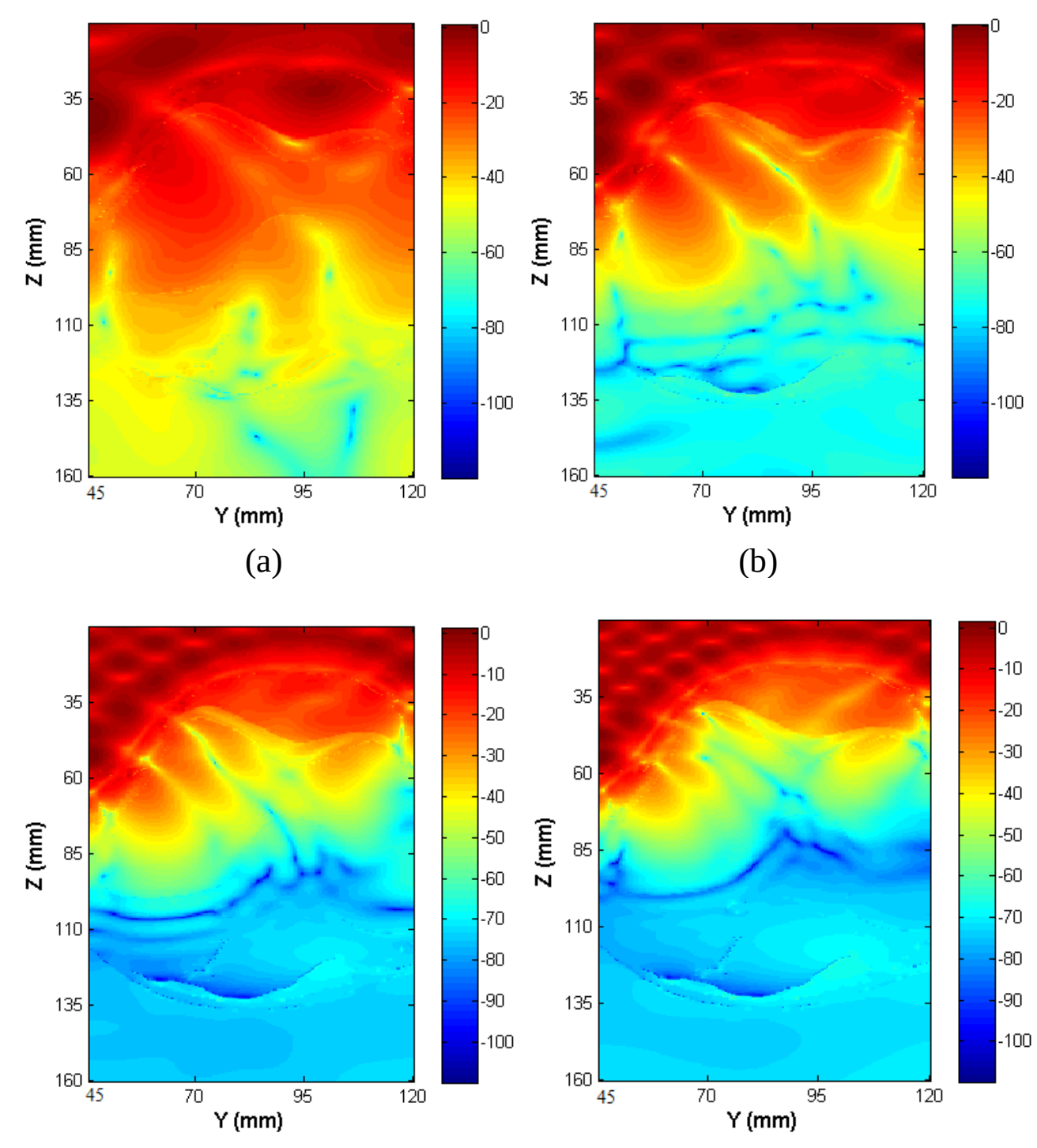

Figure 5 The electric field in the transverse observation plane. (a) 3 GHz; (b) 5 GHz; (c) 7 GHz; (d) 9 GHz.

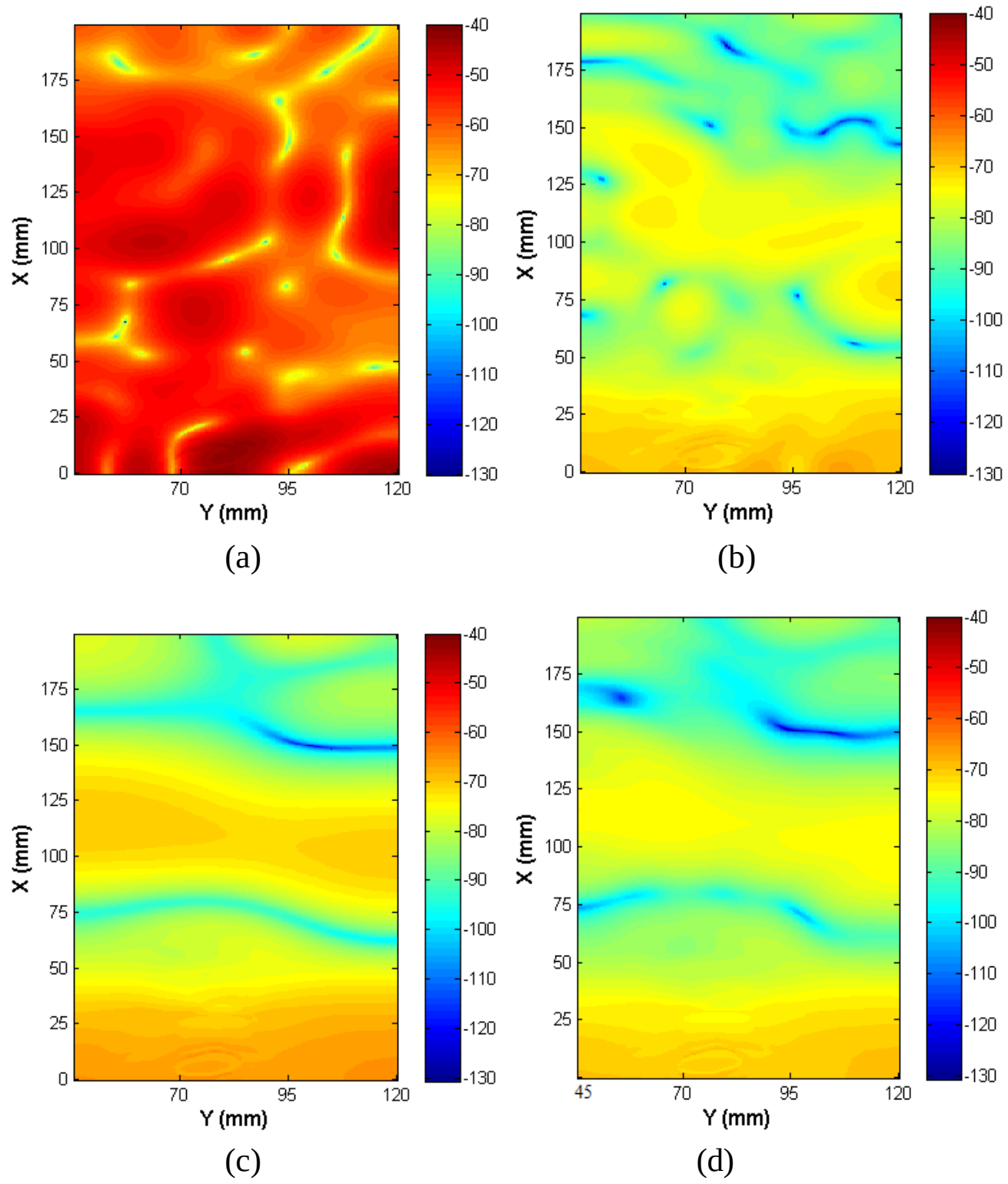

Figure 6. The electric field in the vertical plane behind the knee shown in Figure 3(a). (a) 3 GHz, (b) 5 GHz, (c) 7 GHz, and (d) 9 GHz.

tissue in the center of the knee is bone which has low conductivity value and it changes slightly when the frequency is higher than 6 GHz.

Fig. 6(a-d) shows the electric field at 3, 5, 7, and 9 GHz respectively in the vertical observation plane behind the knee. Again, we noticed when the frequency was higher than 6 GHz, there was no large difference in the distribution of the electric field within this plane. Frequencies higher than 9 GHz were not explored because at which frequency the number of grids for the shortest wavelength in the knee is 19.6 (i.e. $\lambda_{s9GHz} = 19.6\,\Delta, \Delta = 0.25\,mm$). Accuracy cannot be guaranteed if our simulation results are used for higher frequency analysis.

### *B. Time-domain Analysis*

In time-domain UWB band imaging algorithms [35-39], a pulse is radiated to the object and the time delay of the signal during the propagation is estimated and compensated. Therefore, the attenuation and the distortion of the pulse signal due to the object are worthy of investigation.

As shown in Fig. 7, a point source takes turns radiating a short pulse polarized in the X direction from four different positions around the knee: front, behind, left, and right. An observation point was in the center of the knee which is thought to be the deepest place in the knee, since the location of the source and receiver can be moved around the knee. The electric field was recorded at the observation point in the time domain during the entire simulation. The source and the observation point were positioned in the same transverse plane: X=100 *mm*. The short pulse is a modulated Gaussian pulse having a 3 dB bandwidth from 0.1 – 7 GHz while the peak appears at 4.2 GHz. We assumed the model was again immersed in the same coupling medium as in Part A. Since we are interested in the change of the signal's form only due to the knee, for comparison another simulation without the knee was conducted with the source and the observation point located at the same positions. Essentially there is only coupling medium within this object-free model.

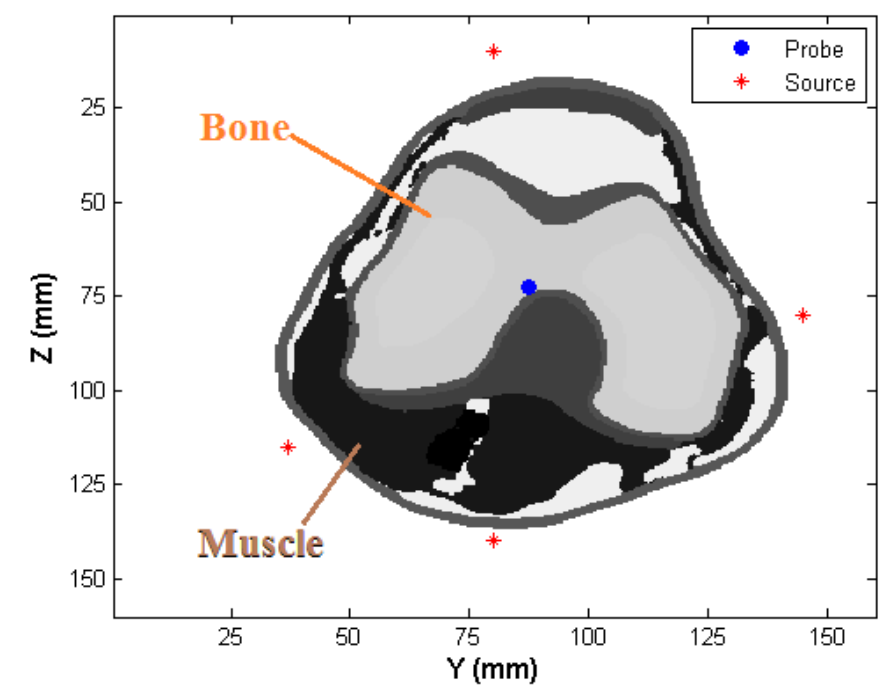

Figure 7. The location of the point-source-illumination to the knee. Red stars indicate the positions of the point source, and the blue dot reprents the position of the observation point.

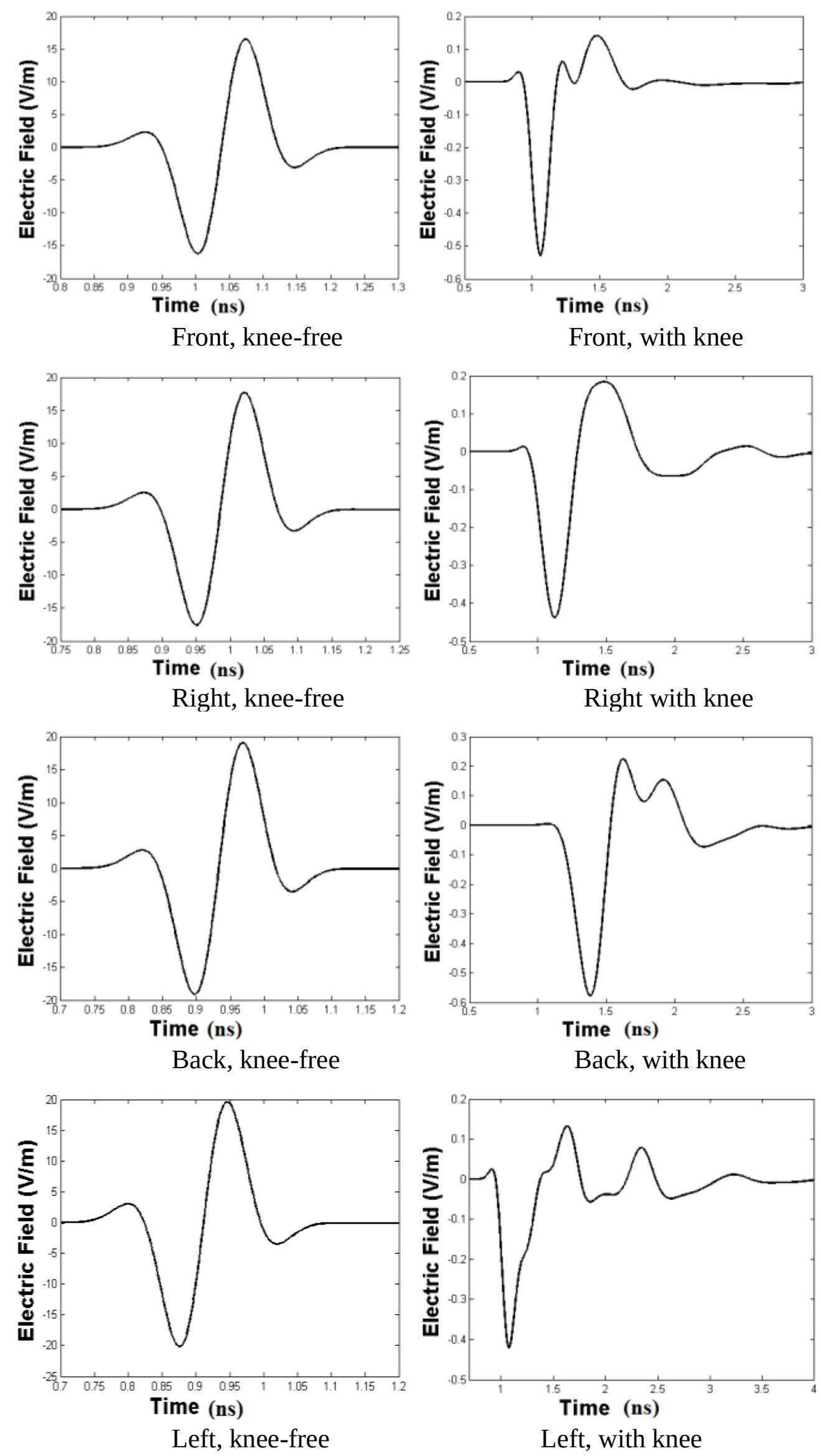

Figure 8. Electric field in the knee's center place with a point source excitation. The left column is without the knee present and the right column is when the knee is present.

Fig. 8 shows the comparison of the electric field at the observation point with and without the knee model. The left column represents the electric-field signal recorded without the knee included, when the source locates in four different positions as shown in Fig 6, respectively. The right column represents the signal recorded at the observation in the knee when the source is located in four positions like before. Comparing the amplitude of the first trough in each row pair in Fig. 8, we found that the attenuation was -29.7 dB, -30 dB, -32 dB, and -33 dB when the source was located in the front, right, back, and left of the knee, respectively. Note that when the source is located in the back and left positions, the signal must pass through the muscle layer, which has high permittivity and conductivity values. The distortion of the signal is also the worst among the four cases when the source is to the left. In reality, probes are positioned outside of the knee, so the attenuation due to the knee would be expected to be double of that found. As such, an attenuation of at least 60 dB should be considered if the receiver is positioned on the other side of the knee (i.e., the signal travels through the knee). If the backscattered signals (for the mono-static mode; or receivers are positioned on the same side as the transmitter in multi-static mode) are used for signal processing, one might need to consider a 120 dB attenuation resulting from the tissue absorption and surface reflection of the knee.

## IV. An Example Of Osteophyte Detection

Knee osteophytes are bony projections that form around the joint margins. They are one of the most common knee diseases and typically can limit range of motion in addition to causing pain. Osteophytes are often associated with arthritis and are a sign of an underlying problem, rather than being a standalone medical issue [40]. A screening test can be used to exam the knee, evaluate the disease, and aid in the formation of a treatment plan.

As a simple example, we consider detecting an osteophyte present in a very common place in the knee using a short microwave pulse. The pulse employed is the one we used in Section III for time domain analysis. The transmitter as well as a receiver array having 25 elements was placed close to the surface of the knee to form a synthetic array [41], as shown in Fig. 9 (not all the receivers are displayed in the figure). Time domain signals collected without the osteophyte will be used for the calibration purpose. By changing the permittivity and conductivity values of cartilage to bone we added in an abnormal coronary growth on the surface of the bone, shown in Fig. 9

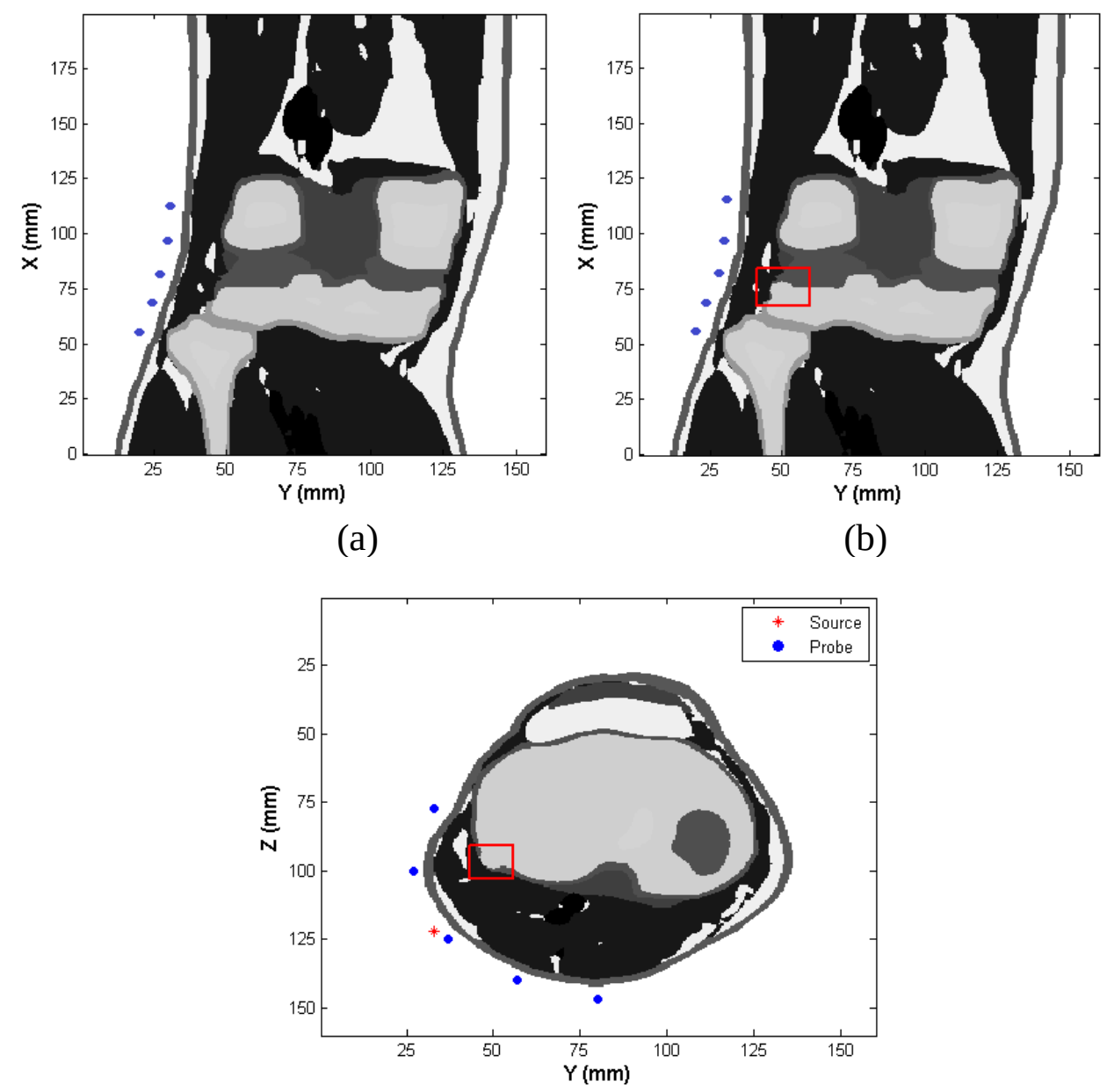

Figure 9. The healthy knee model (a) and the knee model with an osteophyte (b) and (c). The receiver array is composed of 25 elements, in five rows with each row containing 5 elements placed near the osteophyte. Only a part of the receiving elements are shown in the figure. The location of the osteophyte is circled by a red frame to highlight.

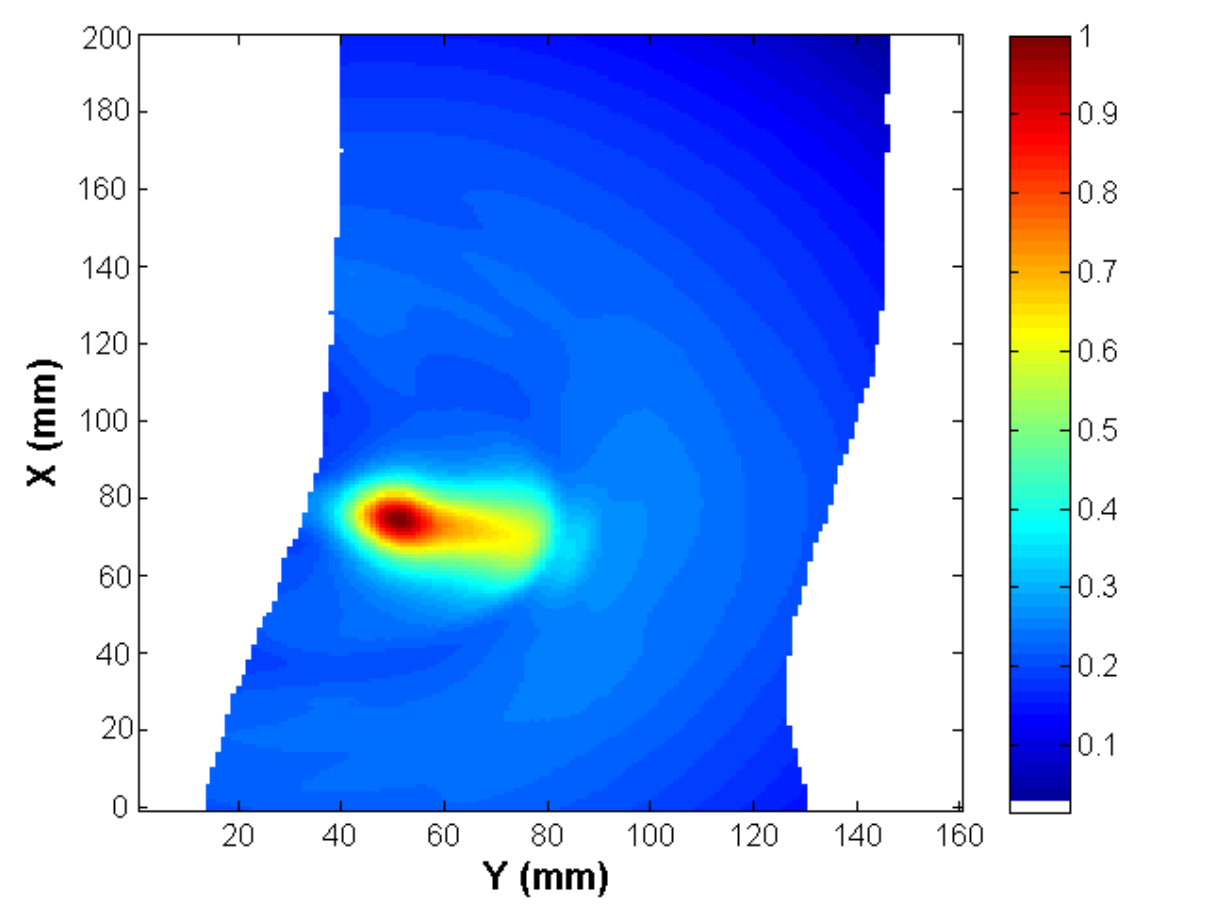

Figure 10. Reconstructed image in the X-Y plane in which the center of the osteophyte locates. The red dot shows the position of the osteophyte.

(b) and (c). The abnormal growth is approximately 1 $cm^2$ big and has maximal thickness 4 mm. By subtraction of two simulations we may obtain the backscattered signal only due to the abnormal growth.

The knee models as well as the antenna array were assumed to be immersed in the coupling liquid as before to reduce impedance mismatch. An eight-core parallel FD-FDTD was used to accelerate the FD-FDTD simulation.

The reconstruction method we used is the one previously developed for breast cancer microwave imaging [36]. This method is a time-domain approach which only aims at discovering and locating any abnormal growth instead of reconstructing the dielectric parameters. This method is fast (an image can be obtained less than 1 minute on an Intel Core i7 computer) and has high robustness (may still work even if the wave form is distorted to some extent, or time delay is not well estimated). The contrast of the reconstructed image is lower when compared to using this method for breast cancer imaging. Therefore, we post processed the image by using MATLAB's inner function "imadjust" to better see the abnormal growth.

The reconstructed image is shown in Fig. 10. Signal intensities outside the knee were set to zero. A dark red area appears in the same position as the osteophyte in the knee; hence our FDTD simulation has provided the correct data for imaging the abnormal growth. This implies the microwave method for detecting an osteophyte in the knee might be feasible. Although not done in this work, one can improve the imaging quality by allowing 25 antennas to take turns sending out the pulse with one antenna serving as a transmitter while the others serve as receivers. This may cause more signals to be involved in signal processing allowing for an improved result. Additionally, increasing the number of antennas (beyond 25) is another option to acquire more signals incorporated into processing to improve results.

## V. Conclusion

Today, while machine-learning approach has been implemented for super-fast EM problem modelling [42], before their accuracy and reliability are well investigated, high-speed EM computation technique is still a good tool for EM-medical problem analysis. In this paper, we developed a 3-D parallel FD-FDTD method to analyze the feasibility of microwave imaging for the diagnosis of the human knee. The attenuation of microwave signals in the knee is not as serious as in the breast. This likely results from tissues in the knee being dominated by relatively low-loss media, such as the bone. This allows for using a higher frequency than previously applied in breast cancer detection possible to acquire a good resolution. However, unlike in breast tumor scanning where tumors are within a relatively uniform background (dominated by fatty tissue, plus glandular tissues), there are more types of tissues in the knee complicating the structure and making the imaging challenging. Further, numerous types of diseases for the knee exist beyond just cancer as typically explored for breast imaging. As such, an advanced image reconstruction algorithm is desired for microwave knee diagnosis and serves for the basis of future studies.